# INSTALLATION STATUS OF THE ELECTRON BEAM PROFILER FOR THE FERMILAB MAIN INJECTOR*

R. Thurman-Keup#, M. Alvarez, J. Fitzgerald, C. Lundberg, P. Prieto, M. Roberts, J. Zagel,
FNAL, Batavia, IL 60510, USA
W. Blokland, ORNL, Oak Ridge, TN 37831, USA

*Abstract*

The planned neutrino program at Fermilab requires large proton beam intensities in excess of 2 MW. Measuring the transverse profiles of these high intensity beams is challenging and often depends on non-invasive techniques. One such technique involves measuring the deflection of a probe beam of electrons with a trajectory perpendicular to the proton beam. A device such as this is already in use at the Spallation Neutron Source at ORNL and the installation of a similar device is underway in the Main Injector at Fermilab. The present installation status of the electron beam profiler for the Main Injector will be discussed together with some simulations and test stand results.

## INTRODUCTION

Traditional techniques for measuring the transverse profile of proton beams typically involve the insertion of a physical object into the path of the proton beam. Flying wires for instance in the case of circulating beams, or secondary emission devices for single pass beamlines. With increasing intensities, these techniques become difficult, if not impossible. A number of alternatives exist including ionization profile monitors, gas fluorescence monitors, and the subject of this report, electron beam profile monitors.

The use of a probe beam of charged particles to determine a charge distribution has been around since at least the early 1970's (see [1] for references to previous devices). The most recent incarnation of this technique is a profile monitor in the accumulator ring at SNS [2].

An Electron Beam Profiler (EBP) has been constructed at Fermilab and has been installed in the Main Injector (MI). The MI is a proton synchrotron that can accelerate protons from 8 GeV to 120 GeV for use by a number of neutrino experiments, and eventually several muon-based experiments. The protons are bunched at 53 MHz with a typical rms bunch length of 1-2 ns. In this report we discuss the design and installation of the EBP and present some studies of the electron beam and simulation results for the anticipated measurement technique.

## THEORY

The principle behind the EBP is electromagnetic deflection of the probe beam by the target beam under study (Fig. 1).

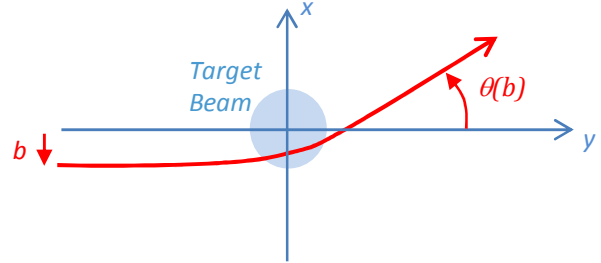

Figure 1: Probe beam deflection (red) for some impact parameter $b$.

If one assumes a target beam with $\gamma \gg 1$, no magnetic field, and $\rho \neq f(z)$, then the force on a probe particle is

$$\vec{F}(\vec{r}) \propto \int d^2\vec{r}' \rho(\vec{r}') \frac{(\vec{r} - \vec{r}')}{|\vec{r} - \vec{r}'|^2}$$

and the change in momentum is

$$\Delta \vec{p} = \int_{-\infty}^{\infty} dt \ \vec{F}(\vec{r}(t))$$

For small deflections, $\vec{r} \approx \{b, vt\}$, and the change in momentum is

$$\Delta \vec{p} \propto \int_{-\infty}^{\infty} dx' \int_{-\infty}^{\infty} dy' \ \rho(x', y')$$
$$\cdot \int_{-\infty}^{\infty} dt \ \frac{\{b - x', vt - y'\}}{(b - x')^2 + (vt - y')^2}$$

where $\{\}$ indicates a vector. For small deflections, $\vec{p} \approx \{0, p\}$ and the deflection is $\theta \approx \frac{|\Delta \vec{p}|}{|\vec{p}|}$. The integral over time can be written as $\text{sgn}(b - x')$ leading to an equation for the deflection

$$\theta(b) \propto \int_{-\infty}^{\infty} dx' \int_{-\infty}^{\infty} dy' \, \rho(x', y') \, \text{sgn}(b - x')$$

where $\text{sgn}(x) = -1$ for $x < 0$ and $+1$ for $x \geq 0$.

If one takes the derivative of $\theta(b)$ with respect to $b$, the sgn function becomes $\delta(b - x')$ leading to

$$\frac{d\theta(b)}{db} \propto \int_{-\infty}^{\infty} dy' \, \rho(b, y')$$

which is the profile of the charge distribution of the beam. Thus for a Gaussian beam, this would be a Gaussian distribution and the original deflection angle would be the error function, $\text{erf}(b)$. This of course is true only to the extent that the above assumptions are valid.

## EXPERIMENTAL PROCEDURE

There are a number of techniques for obtaining $\theta(b)$. A fast scan of the electron beam diagonally through the proton bunch can in principle achieve a measurement in one pass of the bunch. This requires a deflection of the electron beam in a period that is much shorter than the proton bunch. For the MI, this would be sub-nanosecond and may be difficult to achieve.

A second method involves slowly stepping the electron beam through the proton beam and recording a deflection value on each turn of the proton bunch (Fig. 2). In this method the electron beam is stationary each time the proton bunch passes, and then is moved to the next impact parameter.

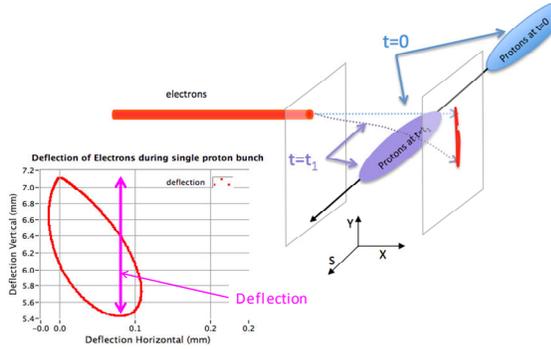

Figure 2: Trajectory followed by a stationary electron beam as the proton bunch passes by. There is some deflection along the proton beam direction due to the magnetic field of the proton beam, but it is much smaller than the deflection transverse to the proton beam.

A variation on the slow scan is to scan quickly along the proton beam direction and slowly transverse to the beam (Fig. 3). The fast scan along the proton beam has a duration similar to the bunch length and allows one to obtain a measurement of the longitudinal beam structure. This longitudinal information can be compared to other instruments and used as a check of the scanning procedure. Additionally, since it is effectively the deflection as a function of longitudinal position within the bunch, it would in principle allow a measurement of the slice profile. At the very least one should be able to obtain head-tail differences.

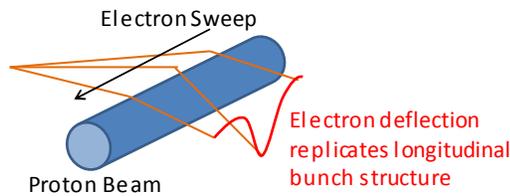

Figure 3: Deflection when the electron beam is scanned along the direction of the proton bunch with a duration similar to the bunch structure.

## APPARATUS

The device (Fig. 4) that was constructed for the MI consists of the EGH-6210 electron gun from Kimball Physics, followed by a cylindrical, parallel-plate electrostatic deflector, and terminating in a phosphor screen.

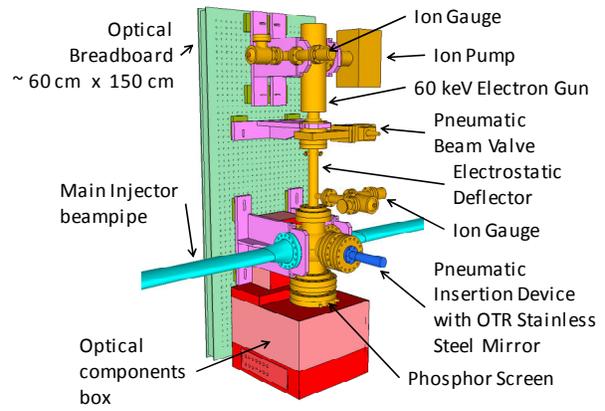

Figure 4: Model of the EBP showing the main components.

The gun (Fig. 5) is a 60 keV, 6 mA, thermionic gun with a $LaB_6$ cathode, that can be gated from 2 μs to DC at a 1 kHz rate. The gun contains a focusing solenoid and four independent magnet poles for steering/focusing. The minimum working spot size is <100 μm. The electrostatic deflector (Fig. 5) contains 4 cylindrical plates that are 15 cm long and separated by ~2.5 cm. Following the electrostatic deflector is the intersection with the proton beamline. There is a pneumatic actuator at this point with a stainless steel mirror for generating optical transition radiation (OTR) to be used in calibrating the electron beam.

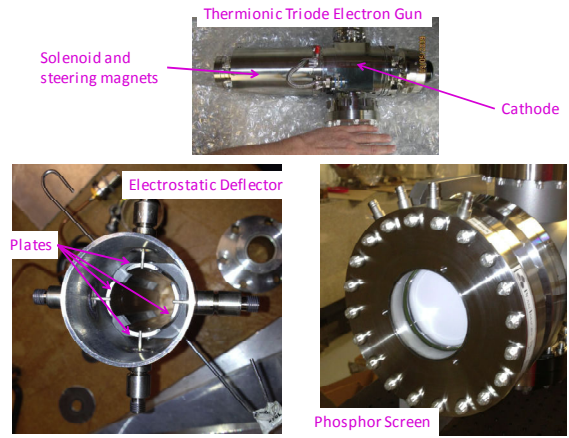

Figure 5: Left) Inside view of the electrostatic deflector showing the cylindrical parallel plates. Right) Phosphor screen mounted in an 8 in conflat flange. A drain wire is attached between the screen and one of the SHV connectors.

After the proton beam intersection there is a phosphor screen from Beam Imaging Systems (Fig. 5). It is composed of P47 ($Y_2SiO_5$:Ce3+) with an emission wavelength of 400 nm, a decay time of ~60 ns and a quantum yield of 0.055 photons/eV/electron. The phosphor screen has a thin conductive coating with a drain wire attached.

Both the OTR and the phosphor screen are imaged by a single intensified camera system (Figs. 6 and 7). The source is chosen by a mirror on a moving stage. Each source traverses a two-lens system plus optional neutral density filters or polarizers before entering the image intensifier (Hamamatsu V6887U-02). The output of the intensifier is imaged by a Megarad CID camera from Thermo-electron (now Thermo Scientific) with a C-mount lens.

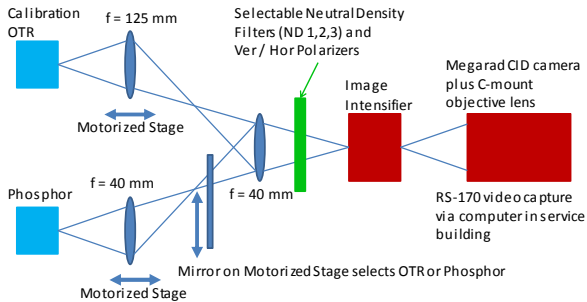

Figure 6: Optical paths followed by the OTR light and the phosphor screen light. Of the two lenses in each path, one is shared.

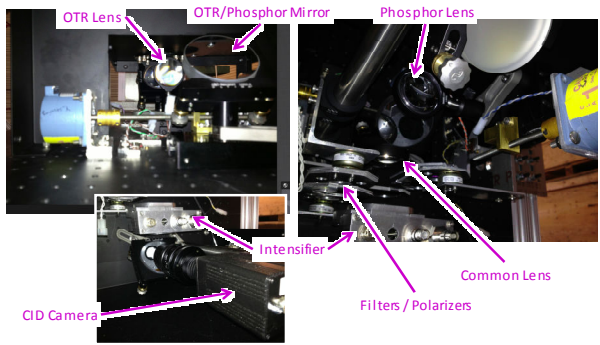

Figure 7: Optical components mounted inside box. The top left picture is looking vertically along the OTR line. The top right picture shows the phosphor path.

## TEST RESULTS

A test stand was setup to measure beam characteristics of the electron gun (Fig. 8). It consisted of a pair of screens used to measure the spot size and divergence to verify the manufacturer's specifications and for use in the simulation.

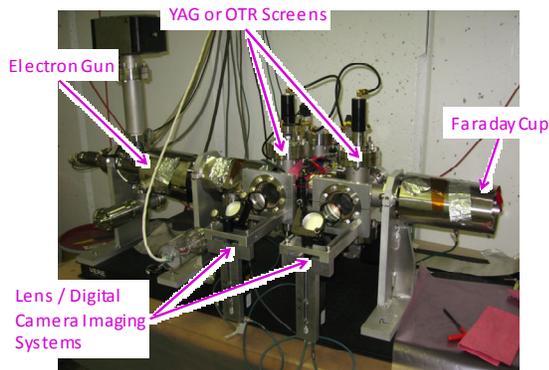

Figure 8: Test stand for measuring beam parameters.

The beam measurements were carried out using the solenoidal magnet in the gun to focus the beam at the first screen, allowing a measurement of the emittance of the electron beam (Fig. 9). Though initial tests were done at 50 keV, the intensity of the MI beam will require an electron energy of around 15 keV.

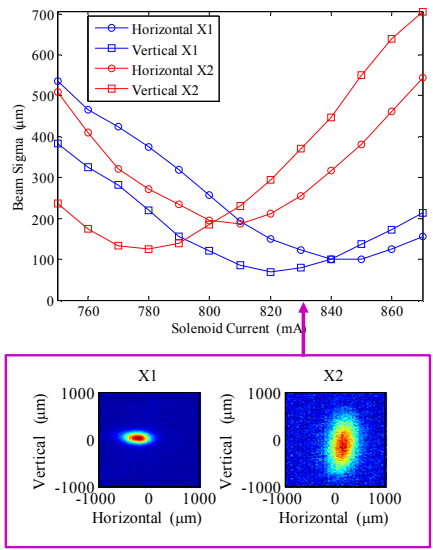

Figure 9: Horizontal and vertical rms beam sizes at the first (blue) and second (red) crosses in the test stand. The measurements are from OTR taken at ~50 keV and 1 mA beam current onto the stainless steel mirrors.

## SIMULATIONS

### Electron Beam

Simulations of the electron beam were developed both at SNS and Fermilab. The SNS calculations (Fig. 10) showed that the measured profile was within 2% of the actual profile. This simulation was based on the slow stepping method and utilized a pencil beam of electrons.

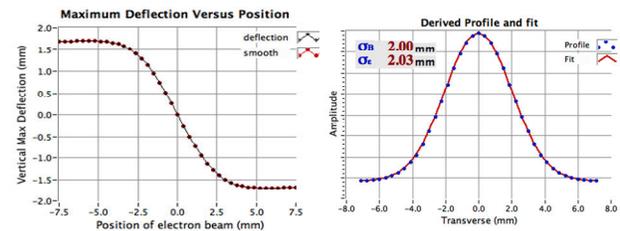

Figure 10: Deflection plot of pencil electron beam, and the derivative of it, showing agreement to better than 2%.

At Fermilab, an electron beam simulation was developed in MATLAB to track electrons through the deflector and proton beam to the phosphor screen. The simulation starts with the measured emittance of the electron gun and propagates the beam through a 2-D calculation of the deflector electric field and through the 3-D electric and magnetic fields of a proton bunch. The time dependence of the deflector field is handled by a linear scaling of the fields. The time dependence of the proton bunch position however, is fully accounted for in evaluating the fields at a given point in time. This simulation was focused on using fast sweeps along the proton direction while slowly stepping through the proton beam. Some results of this simulation are shown in Fig. 11.

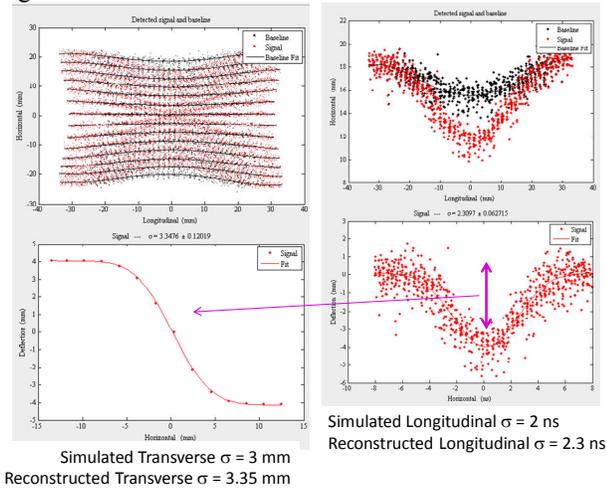

Simulated Transverse σ = 3 mm
Reconstructed Transverse σ = 3.35 mm

Simulated Longitudinal σ = 2 ns
Reconstructed Longitudinal σ = 2.3 ns

Figure 11: Simulated deflection data for varying impact parameters. The black points are baseline deflections with no beam. They result from the non-uniform deflector field. Each point represents a single electron with the random spread given by the measured emittance.

### External Magnetic Fields

External magnetic fields are a serious problem for low energy electron beams. From calculations, a 2 G transverse field will deflect the 15 keV electron beam by 100 mm from gun to phosphor screen. This makes the device inoperable. Figure 17 shows the magnetic fields from the MI magnet busses which are located ~50 cm from the electron beam. The busses run in pairs to mostly cancel the magnetic fields, but there are still low levels remaining. A CST calculation was done to assess the effectiveness of mumetal shielding (Fig. 12).

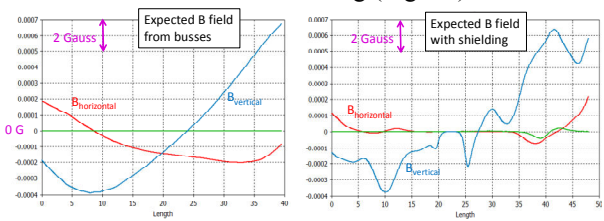

Figure 12: CST simulations of magnetic field from magnet busses along the line of the electron beam. The horizontal component is most important as the electron beam is vertical. Left) The expected dipole field from the magnet busses. Right) The expected field with three layers of mumetal shielding.

The simulation and bench tests both indicated that three layers should be sufficient to eliminate most of the fields. The layers of mumetal were wrapped around the various sections of the EBP separated by welding cloth (Fig. 13). If the mumetal does not work sufficiently, further shielding of the busses may be required.

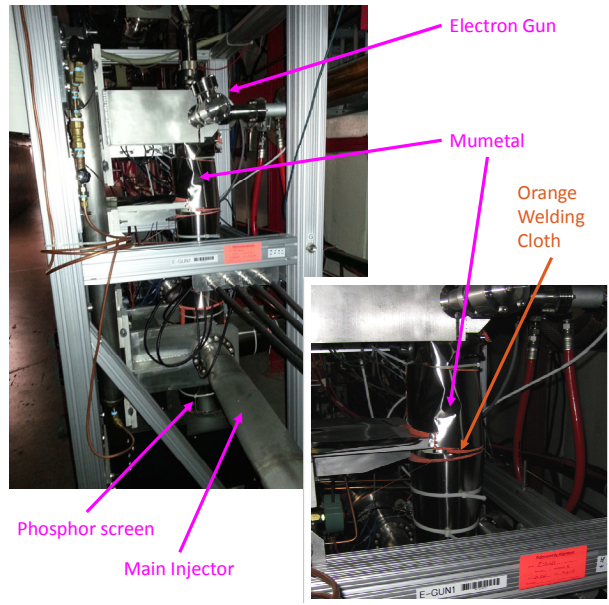

Figure 13: View of the electron beamline showing the mumetal covering.

## INSTALLATION

The EBP was installed in the MI during the 2014 shutdown (Fig. 14). The location is near the end of a straight section just upstream of a horizontal defocussing quadrupole (Q622). The expected horizontal rms beam size at this location is several millimetres.

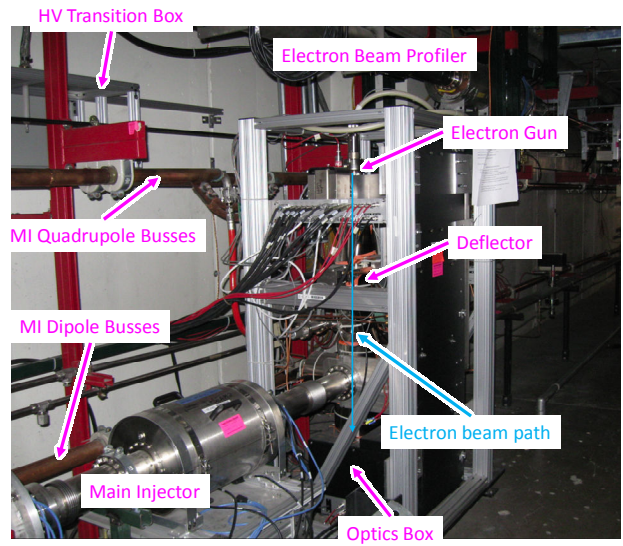

Figure 14: EBP installed in the end of a straight section in the MI. One can see the close proximity to the magnet busses.

Since the installation, effort has been underway to construct the HV transition boxes (Figs. 15 and 16) that connect the commercial cables both from the controller in the service building and the gun in the tunnel, to the RG 220 cables that run from the service building to the tunnel.

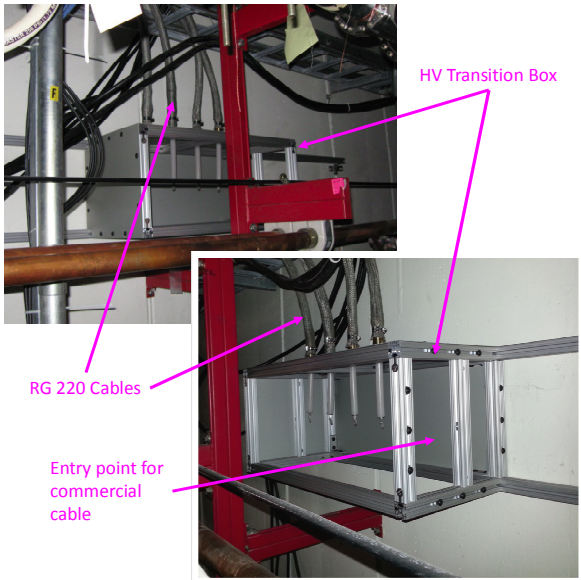

Figure 15: HV transition box in the tunnel. The cables entering from the top are the RG 220 cables. The commercial cable to the gun will exit from the side panel.

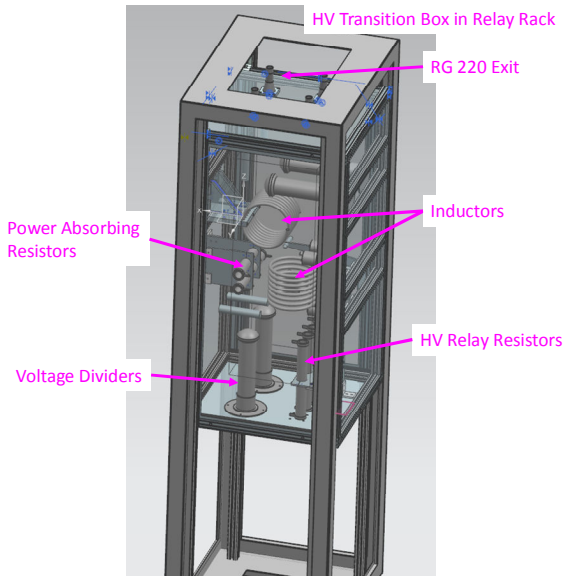

Figure 16: HV Transition box in the service building. This box contains HV relays connected to the safety interlock system, and power absorbers to handle the stored energy in the RG 220 cables.

The choice of RG 220 cables was driven by two facts: the desire to use the commercial controller which limited the resistance of the cables connecting the cathode; and the availability of the RG 220 cables which were left over from the antiproton kicker magnets.

The service building transition box has a number of HV protection and interlock features as seen in the schematic in Fig. 17. Each of the 4 connections to the gun (2 cathode connections, 1 triode gate, 1 HV reference labelled Vgun in the schematic) have an interlock relay to ground for safety. The gate and HV reference connections also have inductors and resistors in series to protect the controller from the stored energy in the ~80 m cables.

The transition box in the tunnel simply makes the connections between the RG 220 and the commercial cable and has no electrical elements.

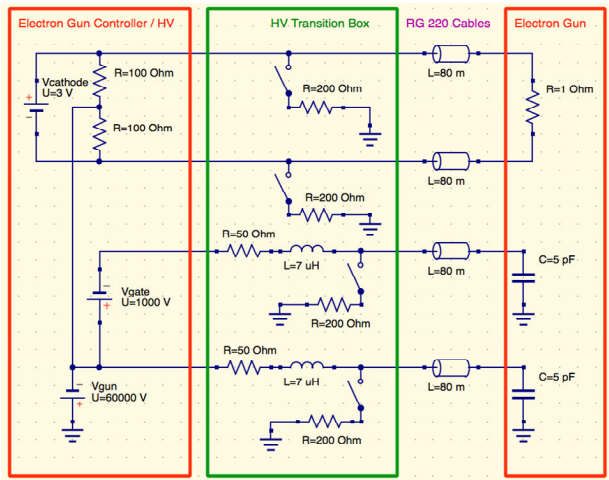

Figure 17: Schematic of HV connections to the electron gun.

## SUMMARY

The EBP has been installed in the MI tunnel with mumetal shielding surrounding the electron beamline. The HV transition boxes are in the process of being assembled with the goal of all work in the tunnel being completed before the end of the current shutdown. Initial turn on of the system is planned for early 2016.

## ACKNOWLEDGMENT

The authors would like to acknowledge the help of the Main Injector, Mechanical Support, Electrical Engineering, and Instrumentation departments for all their assistance in the construction and installation of this device.